\documentclass[11pt,twoside]{article}

\usepackage{asp2006}
\usepackage{epsf}
\usepackage{natbib}

\begin{document}
\title{ThReT: A new survey for Extrasolar Planetary transits at Mt. Holomon, Greece}  
\author{John Antoniadis,$^{1}$ Vassilis Karamanavis,$^{1}$ Dimitris Mislis,$^{2}$ Athanasios Nitsos,$^{1}$ and John H. Seiradakis$^{1}$}   
\affil{$^{1}$Aristotle University, Department of Physics, Section of Astrophysics, Astronomy and Mechanics, $GR-54124$, Thessaloniki, Greece}
\affil{$^{2}$Hamburger Sternwarte, Gojenbergsweg 112, $D-21029$ Hamburg, Germany}

\begin{abstract} 
We present the instrumentation, the target selection method, the data analysis pipeline and the preliminary results of the \textit{Thessaloniki Research for Transits} project (ThReT). \textit{ThReT} is a new project aiming to discover Hot Jupiter Planets, orbiting Sun-like stars. In order to locate the promising spots for observations on the celestial sphere, we produced a sky-map of the Transit Detection Probability by employing data from the Tycho Catalogue and applying several astrophysical and empirical relationships. For the data reduction we used the ThReT pipeline, developed by our team for this specific purpose.
\end{abstract}


\section{Introduction}
ThReT is a new photometric survey initiated at the Aristotle University‘s facilities at Mt. Holomon, Chalkidiki. The latter offers sub-arcsecond seeing (mean value 0.82”) conditions and limited light pollution skies. The employed instruments for the survey are a Celestron11” f/6.3 telescope and a Fingerlakes PL6303E front illuminated CCD camera, mounted on a Synta EQ-6 german mount and guided through a smaller Konus 4” reflector. The achieved pixel scale is 1.32 arcsec and the field of view is 58x33 arcmin. We use no optical filter for our observations.

\section{Target selection-Transit Detection Probability}
We locate the promising spots on the celestial sphere by segmenting  the Tycho catalogue into virtual fields of view of size equal to the field of view of our instrument. We then derive the astrophysical parameters of each main sequence star that our instrument can observe and calculate the overall Transit Detection Probability of each field of view. We finally visualize the results with Transit Probability Maps and locate the promising spots for observation for each season \citep{hlacc}.

\section{Data Reduction}
We have developed the ThReT pipeline to reduce the data of our survey. The pipeline initially corrects the images for bias and dark noise, pixel-to-pixel (flat fielding), and non-linear response of the CCD chip. It then uses DAOPHOT II\citep{st1987} to perform either aperture or PSF photometry, regarding the density of the stars within the field. After deriving the light-curves, the pipeline uses ensemble photometry \citep{ho1992}, the Sys-Rem algorithm \citep{ta2005} and the TFA algorithm \citep{ko2005}, to remove systematic noise. We search for variable stars by employing the Pulsation Index \citep{kj1992} and the PDM algorithm. The light-curves are also scanned for transit-like signals with the BLS algorithm \citep{ko2002}.

\section{Preliminary results}
In order to test the instrumentation and the data reduction pipeline, we observed the field $\alpha : 22:53:40$, $\delta : +44 30 0.0$ for five continuous  nights between 23/07/2008 and 27/07/2008. We refer to that field as $ExoField \#1$. The later has $71.99\%$ probability for transit detection. The exposure time was 120 sec and the time interval between the exposures was 20 sec. With those observational parameters, we reached $<2 \% $ precision photometry for stars between 9 mag and 14 mag.
We have detected several variable candidates (more than 15 so far). We have also detected a transit - like event at the star $\alpha : 22 54 40$,  $\delta : +44 58 48$. The star is $USNO-A2.0 1275-17995584$, its spectral type is K0 but it is unknown if it belongs to the Main-Sequence.

\section{Future work}
We plan to re-observe $ExoField \#1$ in order to further investigate the discovered variable stars and determine their types. We shall also further investigate the suspected planet. If a transit re-occurs then we will undertake follow-up photometric and spectroscopic observations in order to determine the nature of the variability.

\acknowledgements D. Mislis is supported by a PhD scholarship of the DFG Graduiertenkolleg 1351: \textit{Extrasolar Planets and their Host Stars}.

\end{document}